\documentclass[%
 reprint,
amsmath,amssymb,showkeys,
pra,
floatfix,
]{revtex4-2}

\usepackage[linesnumbered,ruled,vlined]{algorithm2e}
\let\oldnl\nl
\newcommand{\nonl}{\renewcommand{\nl}{\let\nl\oldnl}}
\usepackage{braket}
\usepackage{float}
\usepackage{subcaption}
\usepackage{graphicx}
\usepackage{dcolumn}
\usepackage{enumitem}
\usepackage{lipsum}  
\usepackage{bm}
\usepackage{hyperref}
\hypersetup{
    colorlinks=true,
    linkcolor=blue,
    citecolor=magenta
}
\begin{document}

\preprint{APS/123-QED}

\title{Quantum Chemistry Calculations using Energy Derivatives on Quantum Computers}

\author{Utkarsh Azad}
\email{utkarsh.azad@research.iiit.ac.in}
\author{Harjinder Singh}

\affiliation{%
    Center for Computational Natural Sciences and Bioinformatics, International Institute of Information Technology, Hyderabad.
}%

\date{\today}

\begin{abstract}
	Quantum chemistry calculations such as the prediction of molecular properties and modeling of chemical reactions are a few of the critical areas where near-term quantum computers can showcase quantum advantage. We present a method to calculate energy derivatives for both ground state and excited state energies with respect to the parameters of a chemical system based on the framework of the variational quantum eigensolver (VQE). A low-depth implementation of quantum circuits within the hybrid variational paradigm is designed, and their computational costs are analyzed. We showcase the effectiveness of our method by incorporating it in some key quantum chemistry applications of energy derivatives, such as to perform minimum energy configuration search and estimate molecular response properties estimation of H$_2$ molecule, and also to find the transition state of H$_2$ + H $\leftrightarrow$ H + H$_2$ reaction. The obtained results are shown to be in complete agreement with their respective full configuration interaction (FCI) values.
\end{abstract}

\maketitle


\section{\label{sec:intro}Introduction}

Near-term quantum computers, more generally known as noisy intermediate-scale quantum (NISQ) hardware \cite{nisq_preskill}, support hybrid quantum-classical algorithms such as variational quantum eigensolver (VQE) \cite{peruzzo_vqe}, quantum approximate optimization algorithm (QAOA) \cite{2014arXiv1411.4028F}, etc., for solving various computational problems \cite{moll_barkoutsos, PhysRevX.10.021067, PhysRevX.6.031007}. These algorithms use parameterized quantum circuits (PQCs) which consist of quantum gates that depend on classical parameters. To leverage the power of both quantum and classical processors, these algorithms implement a recursive workflow of the following fashion: (i) a highly entangled parameterized quantum state is prepared by the quantum processor for the measurement of expectation values of one or more observables, (ii) a classical processor tries to minimize a function of these expectation values by optimizing classical parameters that control the preparation of the parameterized quantum state.

In the context of quantum chemistry, a variational algorithm like VQE can be used for determining eigenstates and eigenenergies of observables that correspond to the physical/chemical properties of a chemical system or a chemical reaction. Recent work in the field has mostly focused on developing the theory of VQE and VQE-based algorithms for calculations of molecular ground state energies \cite{PhysRevX.6.031007}, excited state energies \cite{higgott_wang_brierley_2019}, molecular vibrations \cite{mcardle_mayorov_shan_benjamin_yuan_2019}, etc. These contributions have been significant in their own respect despite them not being able to provide any advantage over the classical computational chemistry methods, such as density functional theory (DFT) \cite{baseden_tye_2014}, coupled cluster (CC) theory \cite{aszabo82:qchem}, and quantum Monte-Carlo methods \cite{0f8aef7d72424b6eada5de3cd9269a43}. This lack of advantage is attributed to the fact that much of the work done in the field is still in the exploratory phase, and the computational power offered by NISQ devices is considerably restricted due to the limited number of good quality qubits, absence of error correction and limited qubit connectivity \cite{2021arXiv210401992S}. 

While analyzing molecules, the molecular energy derivatives with respect to some system parameters prove to be as crucial as molecular energies to calculate a range of time-independent physical and chemical properties. For example, (i) first-order derivatives of energy with respect to geometric coordinates allows us to search minimum energy configuration and reaction paths, (ii) higher-order derivatives of energy with respect to external electric fields allows us to predict some key molecular response properties such as (hyper)polarizability, magnetizability, etc. This work presents a VQE-based method to calculate molecular energy derivatives on a quantum computer for both ground state energy and excited state energies up to the second order. Low-depth circuit implementations for our methods are designed and their feasibility on the NISQ hardware is analyzed. We show the use of these energy derivatives for the following quantum chemistry tasks: (i) minimum energy configuration search for H$_2$ molecule, (ii) estimation of molecular response properties such as dipole moment and polarizability for H$_2$ molecule, and (iii) transition state search for the reaction H$_2$ + H $\leftrightarrow$ H + H$_2$. Our variational method gives results in complete agreement with those obtained using full configuration interaction (FCI) values.

\textit{Structure}: In Section \ref{sec:vqe}, we introduce the framework of variational quantum eigensolver. The methodology for finding energy derivatives is described in Section \ref{sec:energy-derivative} for both ground state energy and excited state energies. Then, in Section \ref{sec:results}, we showcase our results for the quantum chemistry tasks as mentioned earlier. Finally, we conclude with a discussion and overview of possible improvements in Section \ref{sec:discussion-conclusion}.

\vfill{}

\begin{figure*}
    \centering
    \includegraphics[width=\linewidth]{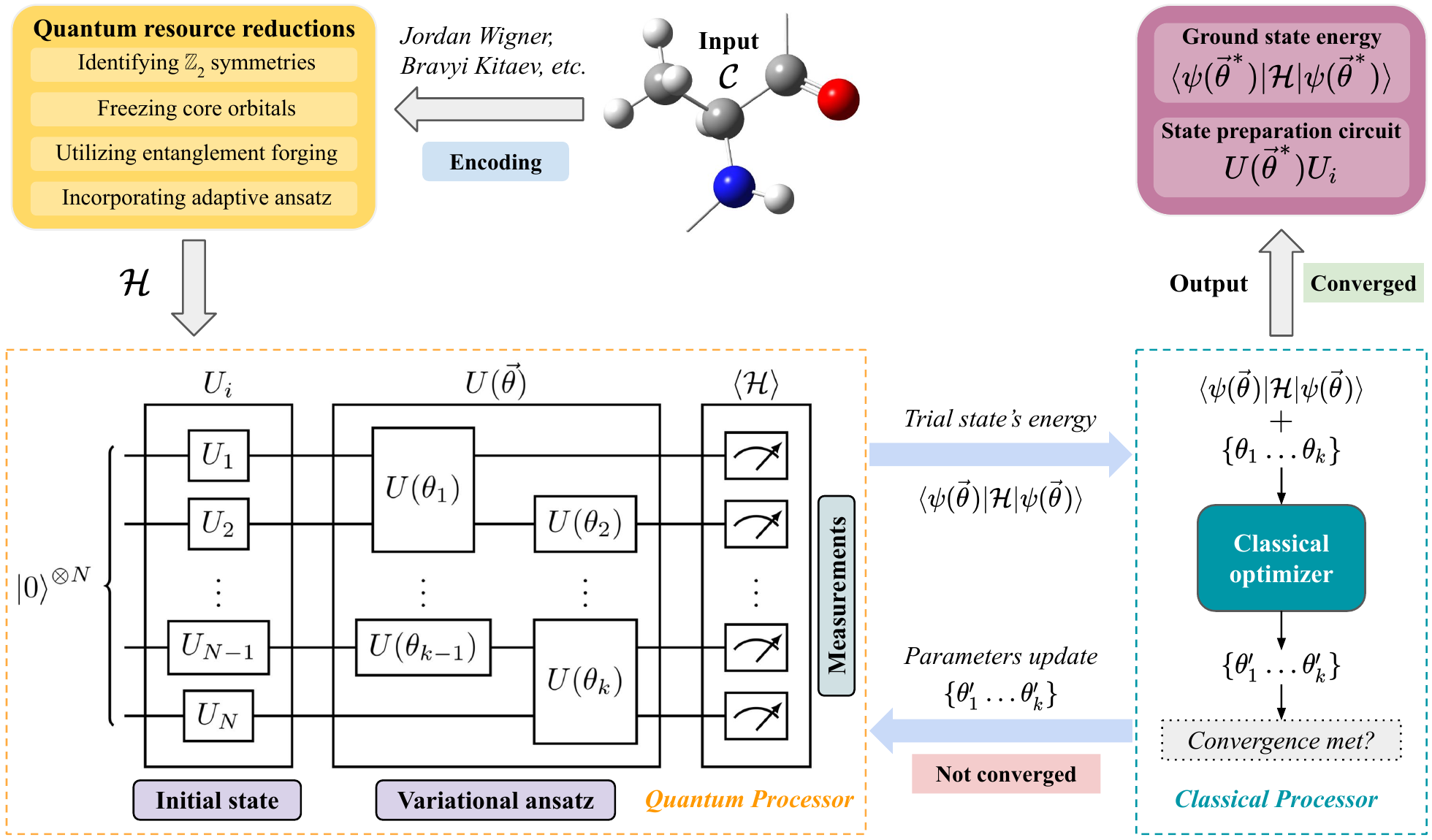}
    \caption{\textbf{VQE Workflow}: Workflow for variational quantum eigensolver (VQE)}
    \label{fig:vqe-workflow}
\end{figure*}

\section{\label{sec:vqe}Variational Quantum Eigensolver}

Variational quantum eigensolver (VQE) is one of the first hybrid quantum-classical algorithms that was proposed as a candidate algorithm for near-term quantum devices by \citet{peruzzo_vqe}. It makes use of a parameterized quantum circuit (PQC), also known as an ansatz \cite{2021arXiv210308505F}, to evolve a known initial trial state $\ket{\psi_{i}}$ to a target trial state $\ket{\psi_{t}}$ and find the eigenstates and eigenvalues for a system or a problem that can be encoded in the Hermitian observable $\mathcal{M}$. For a chemical system $\mathcal{C}$, this procedure (Fig. \ref{fig:vqe-workflow}) is briefly described as follows:

\begin{enumerate}
    \item To encode any property of a $\mathcal{C}$ into a hermitian observable $\mathcal{M}$, one generates corresponding fermionic operators from the relevant one and two-body integrals $\{h_{ij}, h_{ijkl}\}$ in the spin basis obtained using any computational chemistry package like Gaussian \cite{g16}, PySCF \cite{pyscf}, Psi4 \cite{psi4}, etc.
    \begin{equation}\label{eq:1}
        \mathcal{M} = \sum_{ij}h_{ij}a^{\dagger}_{i}a_{j} + \sum_{ijkl}h_{ijkl}a^{\dagger}_{i}a^{\dagger}_{j}a_{k}a_{l}
    \end{equation}
    \item The fermionic operators can be realized as qubit operators by expressing them in terms of Pauli operators using certain encoding schemes such as Jordan-Wigner, Bravyi-Kitaev, etc \cite{tranter_love_mintert_coveney_2018}. We refer to this as the qubitization of $\mathcal{M}$:
    \begin{equation}\label{eq:2}
        \mathcal{M} = \sum_{P\in \mathcal{P}} h_{P} P
    \end{equation}
    where, $\mathcal{P} = \{I,X,Y,Z\}^{\otimes N}$, and $h_{P} \in \mathbb{R}$. The number of qubits $N$ required to encode this is equivalent to number of spin orbitals associated with $\mathcal{C}$ having spin-up ($\alpha$) and  spin-down ($\beta$) electrons. At this stage, one can further analyze $\mathbb{Z}_2$ symmetries \cite{2017arXiv170108213B} present in the chemical system and freeze core orbitals to taper off some qubits.
    \item For the qubitized $\mathcal{M}$, we construct an anstaz, i.e., a parameterized unitary, represented by $U(\vec{\theta})$. In principle, depending on the structure, there are three types of ansatz - (i) hardware-efficient ansatz \cite{2019arXiv191009694R}, (ii) physically-inspired ansatz \cite{RevModPhys.92.015003}, and (iii) adaptive ansatz \cite{2019NatCo.10.3007G}. All of them will evolve the initial trial state $\ket{\psi_{i}}$ as follows:
    \begin{equation}\label{eq:3}
        \ket{\psi_t} \equiv \ket{\psi(\vec{\theta})} = U(\vec{\theta})\ket{\psi_{i}} \equiv U(\vec{\theta})U_{i}\ket{0}^{\otimes N}
    \end{equation}
    In our case, the preferred initial trial state  $\ket{\psi_{i}} = \ket{\psi_\text{HCF}}$ which is the qubit Hartree-Fock state is prepared using state initialization unitary $U_{i}$ on $\ket{0}^{\otimes N}$. The ansatz $U(\vec{\theta})$ is applied on this state, where the vector $\vec{\theta} = (\theta_1, \theta_2, \ldots, \theta_k)$ has $k$ classical parameters that are provided by a classical processor. The target trial state $\ket{\psi_t} = \ket{\psi(\vec{\theta})}$ should ideally correspond to the FCI or UCCSD ground state wavefunction.
    \item The parameters $\vec{\theta}$ are optimized by a classical processor using Rayleigh–Ritz variational principle \cite{2018arXiv181208767Y} on the expectation value of $\mathcal{M}$ obtained from the quantum processor. For the optimal value of $\vec{\theta} = \vec{\theta}^{*}$, we get:
    \begin{equation}\label{eq:4}
        E(\vec{\theta}^{*}) = \min{\bra{\psi(\vec{\theta})}\mathcal{M}\ket{\psi(\vec{\theta})}} \quad E(\vec{\theta}^{*}) \geq E_{GS}
    \end{equation}
    Sometimes, one may not directly use the expectation value of $\mathcal{M}$ as it is, and rather benefit from performing an additional classical post-processing step. This allows us to shift certain computation onto the classical processor to either reduce amount of quantum resources required as in the case of entanglement forging \cite{2021arXiv210410220E}, or to perform some complex tasks such as building quantum classifiers \cite{Schuld_2017}.
\end{enumerate}

To obtain the ground state of the chemical system $\mathcal{C}$, the $\mathcal{M}$ corresponds to Hamiltonian $\mathcal{H}$ of the system. Similarly, the other properties such as dipole moment, angular momentum, etc, can be calculated via the expectation value of their corresponding qubitized observable with respect to the ground state wavefunction prepared by optimized ground state preparation unitary $U(\vec{\theta}^{*})U_{i}$.

\section{\label{sec:energy-derivative}Energy Derivatives}

The final result of the traditional variational quantum eigensolver (VQE) algorithm is the molecular ground state energy $E_0$ and the corresponding state preparation circuit $U(\vec{\theta}^{*})U_{i}$ for the ground state ($\ket{\psi}_\text{GS}$) itself. However, even for VQE, the molecular energy derivatives $\partial E$ with respect to some parameters are also important quantities that are used for its training and also for extending its capabilities. For example, in the case of training VQE, we calculate $\partial_{\theta_i} E$ for updating the parameterized unitary parameters $\vec{\theta}$ using a gradient-based rule. Similarly, by calculating $\partial_{r_i} E$, i.e., force dependent on nuclear coordinates $\vec{r}$, one can extend VQE to perform geometric structure optimization, transition-state search, etc. Therefore, this makes the task of calculation of energy derivatives crucial and necessary.

In traditional quantum chemistry literature, there exist two ways to calculate these energy derivates $\partial_{\eta} E$. The first one is to use the finite-difference method \cite{10.5555/19316}, and the second is to use analytical formula-based methods such as sum-over-state approach \cite{obrien_2019}. In this section, we discuss the calculation of $\partial_{\eta} E$ using VQE-based strategy.

\subsection{\label{subsec:param-within}Derivatives with respect to \texorpdfstring{$\vec{\theta}$}{0}}

For estimating the derivatives of energy with respect to the variational parameters of the circuit, one can use parameter-shift rules \cite{2021PhRvA.103a2405M}. It involves running the same parameterized circuit with different shifts $s$ in the parameters $\theta_i$.

\begin{equation}\label{eq:5}
    \partial_{\theta_j} E \equiv \frac{\partial \langle \mathcal{H} \rangle}{\partial \theta_{j}} = \frac{\langle \mathcal{H} \rangle_{\theta_j + s} - \langle \mathcal{H} \rangle_{\theta_j-s}}{2\sin{(s)}}
\end{equation}

This makes it possible to implement gradient-based update rules for parameter $\theta_j$ with some learning rate constant $\gamma > 0$:

\begin{equation}\label{eq:6}
    \theta_j^{t+1} = \theta_j^{t} - \gamma \partial_{\theta_j} E
\end{equation}

One can easily extend this to calculate higher-order gradients using a scheme similar to the finite-difference method, i.e., evaluation using a finite shift in the parameter \cite{2021PhRvA.103a2405M}. For example, the calculation of Hessian $\partial_{\theta_i\theta_j} E$ would require the shift of two parameters $\theta_i$ and $\theta_j$ simultaneously by $s_1$ and $s_2$ respectively. 

\subsection{\label{subsec:param-beyond}Parameterising E beyond \texorpdfstring{$\vec{\theta}$}{0}}
In general, the hermitian observable $\mathcal{M}$ described in eq. \ref{eq:1} is dependent upon the basis set taken into consideration, the geometry of the molecule, and the environment, i.e., the effect of the external electric field it is placed in. Since we have limited number of qubits $N$, we would like to focus upon the latter two (geometry and environment), as tweaking the former beyond the minimal basis set will lead to an undesirable increase in $N$. 

Let this dependency on the system parameter other than the variational parameter $\vec{\theta}$ be given by $\vec{\eta} = (\eta_{0}, \eta_{1}, \ldots, \eta_{k})$. Then, $\mathcal{H}$ and $\langle \mathcal{H} \rangle$ are written as:

    \begin{equation}\label{eq:7}
        \mathcal{H}(\vec{\eta}) = \sum_{ij}h_{ij}(\vec{\eta})a^{\dagger}_{i}a_{j} + \sum_{ijkl}h_{ijkl}(\vec{\eta})a^{\dagger}_{i}a^{\dagger}_{j}a_{k}a_{l}
    \end{equation}
    \begin{equation}\label{eq:8}
        \mathcal{H}(\vec{\eta}) = \sum_{P\in \mathcal{P}} h_{P}(\vec{\eta}) P 
    \end{equation}
    \begin{equation}\label{eq:9}
        E(\vec{\theta}^{*}(\vec{\eta}), \vec{\eta}) = \min{\bra{\psi(\vec{\theta})}\mathcal{H}(\vec{\eta})\ket{\psi(\vec{\theta})}}
    \end{equation}
    
As discussed earlier, finding the derivatives $\partial_\eta E$ is essential for studying a molecule and its molecular properties. We describe the methodology for their calculations in subsequent sections using the following shorthand:

    \begin{equation}\label{eq:10}
        E^*(\vec{\eta}) = E(\vec{\theta}^{*}(\vec{\eta}), \vec{\eta})
    \end{equation}

\subsection{\label{subsec:first-order}First-order derivatives: }
To calculate first-order energy derivatives $\partial_\eta E$, we use Hellmann–Feynman theorem \cite{doi:10.1063/1.3266959}, which gives us:
\begin{equation}\label{eq:11}
\begin{split}
        &\frac{\partial E^*(\vec{\eta})}{\partial \eta_i} =\frac{\partial \bra{\psi(\vec{\theta}^{*})}\mathcal{H}(\vec{\eta})\ket{\psi(\vec{\theta}^{*})}}{\partial \eta_{i}} =\\&\underbrace{\frac{\partial \bra{\psi(\vec{\theta}^{*})}}{\partial \eta_i}\mathcal{H}(\vec{\eta})\ket{\psi(\vec{\theta}^{*})}}_\text{$\neq$ 0} + \bra{\psi(\vec{\theta}^{*})}\frac{\partial \mathcal{H}(\vec{\eta})}{\partial \eta_i}\ket{\psi(\vec{\theta}^{*})} +\\ &\underbrace{\bra{\psi(\vec{\theta}^{*})}\mathcal{H}(\vec{\eta})\frac{\partial\ket{ \psi(\vec{\theta}^{*})}}{\partial \eta_i}}_\text{$\neq$ 0}
\end{split}    
\end{equation}

The first and third terms are opposite in sign but $\neq 0$ because the optimal variational parameters $\vec{\theta}^{*}$ are also dependent upon system parameters $\vec{\eta}$. This leads to the following result from eq. \ref{eq:11}:

\begin{equation}\label{eq:12}
    \frac{\partial E^*(\vec{\eta})}{\partial \eta_i} = \bra{\psi(\vec{\theta}^{*})}\frac{\partial \mathcal{H}(\vec{\eta})}{\partial \eta_i}\ket{\psi(\vec{\theta}^{*})}
\end{equation}

The first-order derivatives $\partial_{\eta_i} \mathcal{H}(\vec{\eta})$ can be calculated on a classical processor either analytically by solving coupled perturbed Hartree-Fock equations \cite{doi:10.1063/1.4908131}, or using the finite-difference method \cite{pulay_2013}. \citet{higgott_wang_brierley_2019} showed that the number of measurements $m_j$ required for estimating the expectation value of a subterm $P_j$ of the qubitized Hamiltonian $\mathcal{H}=\sum_P h_P P$ with the variance $\epsilon^{2}$ is $O(|h_P|^2/\epsilon^{2})$. Since, the decomposition of molecular Hamiltonian (eq. \ref{eq:1}) into qubitized Hamiltonian (eq. \ref{eq:2}) results in $O(n^4)$ terms, the overall cost for calculating $\partial E_{\eta}$ with variance $\epsilon^2$ using eq. \ref{eq:12} is $O(n^4 N_{\eta} (\sum_P |h_P|)^2/\epsilon^2)$, where $N_{\eta}$ are the number of system parameters $\vec{\eta}$.

\subsection{\label{subsec:second-order}Second-order derivatives}

Similar to first-order energy derivatives, we can calculate the second-order energy derivatives $\partial^2_\eta E$ as follows: 

\begin{equation}\label{eq:14}
\begin{split}
    \frac{\partial^2 E^*(\vec{\eta})}{\partial \eta_{i}\eta_{j}} =& \frac{\partial}{\partial \eta_{i}}  \bra{\psi(\vec{\theta}^{*})}\frac{\partial \mathcal{H}(\vec{\eta})}{\partial \eta_j}\ket{\psi(\vec{\theta}^{*})} \\=&  \underbrace{\bra{\psi(\vec{\theta}^{*})}\frac{\partial \mathcal{H}(\vec{\eta})}{\partial \eta_i\partial \eta_j}\ket{\psi(\vec{\theta}^{*})}}_\text{I} +  \\& \underbrace{2 \textbf{Re}\Bigg[\bra{\psi(\vec{\theta}^{*})}\frac{\partial \mathcal{H}(\vec{\eta})}{\partial \eta_{j}}\frac{\partial \ket{\psi(\vec{\theta}^{*})}}{\partial \eta_{i}} \Bigg]}_\text{J}
\end{split}    
\end{equation}

\begin{figure}[t]
    \centering
    \begin{subfigure}[b]{\linewidth}
    \begin{minipage}{.05\textwidth}
        \caption{}
        \label{fig:vqe-circuit-full}
    \end{minipage}%
    \begin{minipage}{0.95\textwidth}
        \includegraphics[width=.95\linewidth]{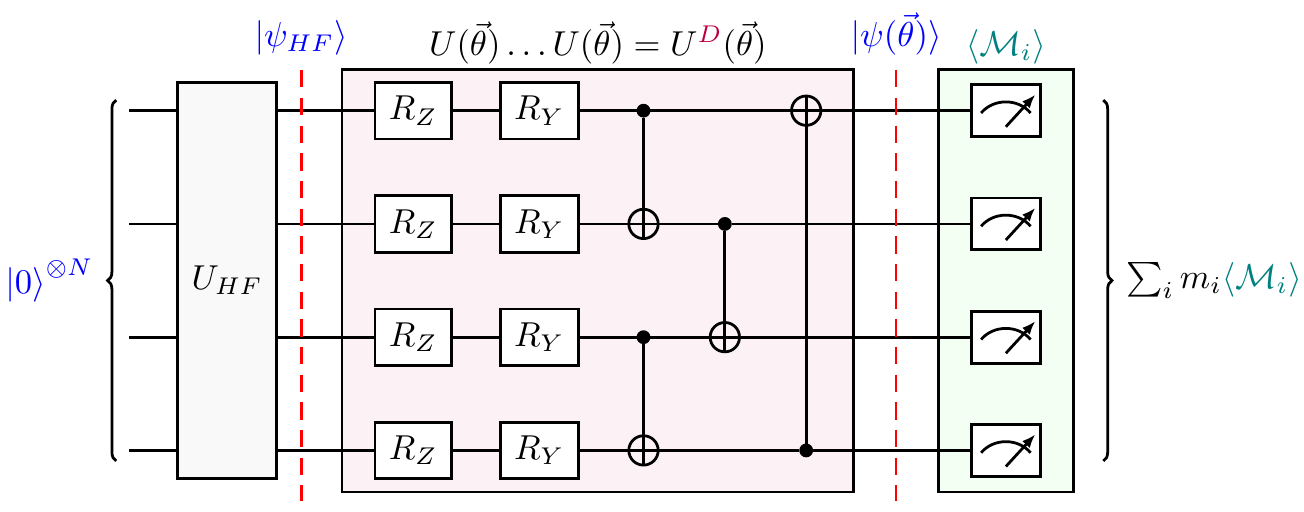}
    \end{minipage}
    \end{subfigure}
    \newline%
    \begin{subfigure}[b]{\linewidth}
    \begin{minipage}{.05\textwidth}
        \caption{}
        \label{fig:vqe-circuit-tapered}
    \end{minipage}%
    \begin{minipage}{0.95\textwidth}
        \includegraphics[width=.95\linewidth]{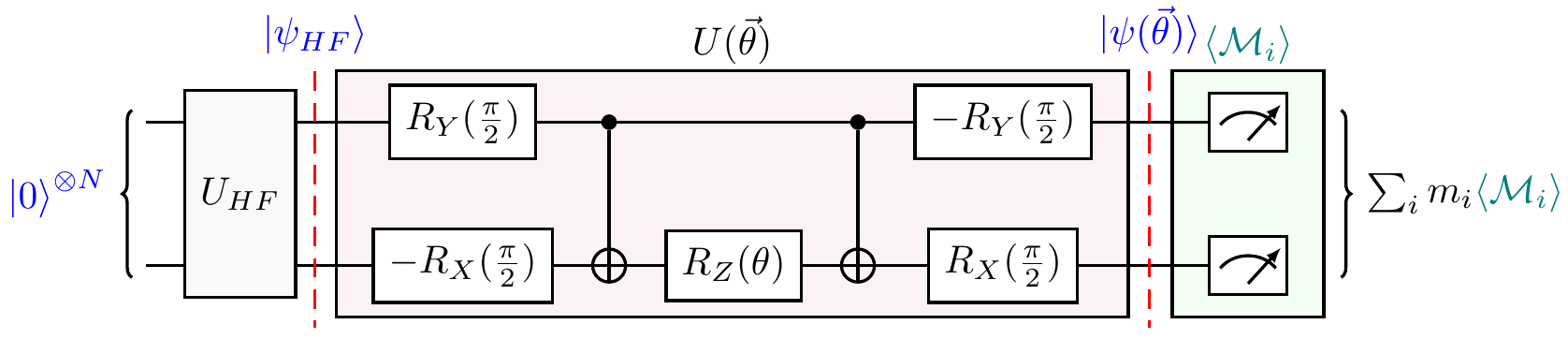}
    \end{minipage}
    \end{subfigure}
    \caption{\textbf{VQE Circuits}: The $U_\text{HF}$ unitary block (gray) evolves $\ket{0}^{\otimes N}$ to the Hartree-Fock state $\ket{\psi_\text{HF}}$. This state is evolved by an ansatz, i.e., a parameterized unitary $U^{D}(\vec{\theta})$ to $\ket{\psi(\vec{\theta})}$, where $D$ is the number of ansatz block (red) is repeated. Finally, Pauli basis measurements are performed on all qubits to calculate expectation value $\mathcal{M}_i = \sum (-1)^{\sum_j x_j} P(x)$, where $x$ is a bit string representation  of possible $2^N$ outcomes.} 
    \label{fig:vqe-circuit}
\end{figure}

The calculation of the first term $I$, i.e., second-order derivative $\partial_{\eta_{i}} \partial_{\eta_{j}} \mathcal{H}(\vec{\eta})$ is again done on a classical processor just as in the case of $\partial_{\eta_i} \mathcal{H}(\vec{\eta})$. However, calculation of the second term $J$ is not so straightforward. We first focus on the term $\partial_{\eta_i}\ket{\psi(\vec{\theta}^*)}$ and expand it as follows:

\begin{equation}\label{eq:15}
   \frac{\partial \ket{\psi(\vec{\theta}^*)}}{\partial \eta_i} = \bigg| \frac{ \ket{\psi(\vec{\theta}^*)_{\eta_i+d\eta_i}} - \ket{\psi(\vec{\theta}^*)}}{d\eta_i} \bigg|_{d\eta_i \rightarrow 0}
\end{equation}

This allows us to rewrite $J$ as the difference of the following two expectation values:

\begin{equation}\label{eq:16}
\begin{split}
    J \equiv \frac{2}{d\eta_i}\textbf{Re}\Bigg[ &\underbrace{\bra{\psi(\vec{\theta}^{*})}\frac{\partial \mathcal{H}(\vec{\eta})}{\partial \eta_{j}} \ket{\psi(\vec{\theta}^*)_{\eta_i+d\eta_i}}}_{J_1} -\\ &\underbrace{\bra{\psi(\vec{\theta}^{*})}\frac{\partial \mathcal{H}(\vec{\eta})}{\partial \eta_{j}} \ket{\psi(\vec{\theta}^{*})}}_{J_2} \Bigg]
\end{split}    
\end{equation}

While the second term $J_2$ has already been calculated during the first-order energy derivative calculations, the first term $J_1$ can be calculated from a low-depth implementation of overlap estimation method proposed in \cite{harrow_kandala_gambetta_2019}. Assume two unitaries $U_1$ and $U_2$ such that (i) $\ket{\psi(\vec{\theta}^{*})} = U_1 \ket{0}^{\otimes N}$, and (ii) $\partial_{\eta_j} \mathcal{H}\ket{\psi(\vec{\theta}^{*})}_{\eta_i+d\eta_i} = U_2 \ket{0}^{\otimes N}$. Then the state preparation circuit $U_1^{\dagger}U_2\ket{0}^{\otimes N}$, i.e. unitary $U_2$ followed by the inverse of unitary $U_1$, can be used for estimating the overlap with variance $\epsilon^\prime$ by measuring the probability of obtaining $\ket{0}^{N}$ state using $O(1/\epsilon^{\prime2})$ shots.

Unlike previously, the calculation of $\partial^2 E$ with variance $\epsilon^2$ using eq. \ref{eq:14} requires $O(n^4 N_{\eta}^2 (\sum_P |h_P|)^2/\epsilon^2)$ measurements. 

\begin{figure}[tp]
    \centering
    \includegraphics[width=\linewidth]{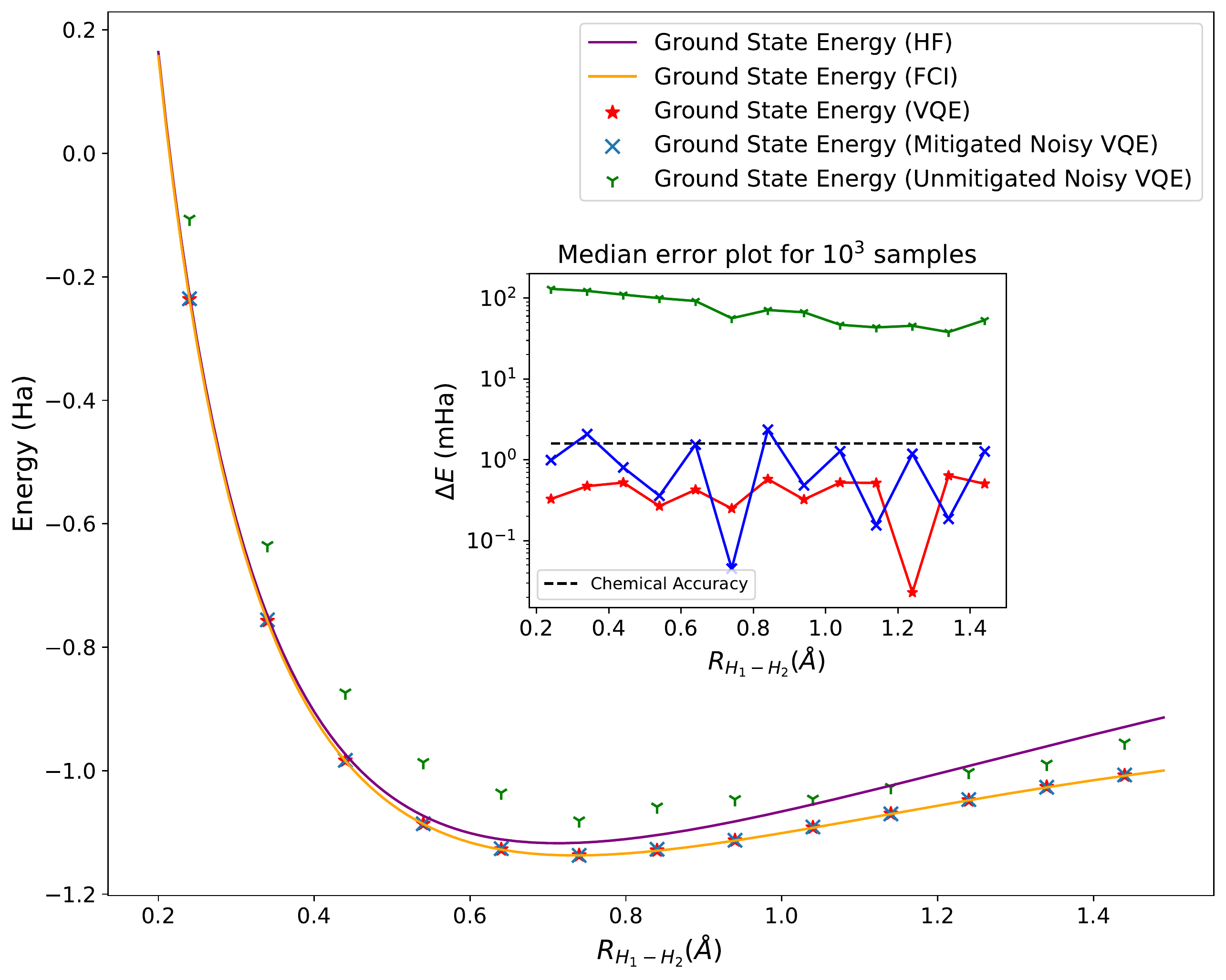}
    \caption{\textbf{VQE Energies for H$_2$ molecule}: Ground state energies computed after varying the bond length $R_{\text{H}_1-\text{H}_2} \in$ [0.2, 1.5] \AA. The solid curve indicates classically computed energy values for Hartree Fock (HF) and full configuration interaction (FCI) methods. Inset represents absolute errors in energies at every bond length $R_{\text{H}_1-\text{H}_2}$ with chemical accuracy (1.6 mHa) taken as reference.}
    \label{fig:vqe-energy}
\end{figure}

\begin{algorithm}[t]
	\SetAlgoLined
	\DontPrintSemicolon
	\SetKwFor{Loop}{loop}{}{end loop}
	\KwData{chemical system $\mathcal{C}$, generalized coordinates $Q$, learning rate $\gamma$, optimizer $opt$, tolerance $ctol$} 
	\textbf{Initialize:} $\mathcal{C}$.molecule $\leftarrow$ (Initial coordinates) \\
	\nonl \hspace{45pt} $\mathcal{C}$.optimized $\leftarrow$ False\\
	\nonl \hspace{45pt} $\vec{\theta}$ $\leftarrow$ (Initial values)\\
	\nonl \hspace{45pt} delQ $\leftarrow$ $[0, \ldots, 0]$,\quad len(delQ) = len(Q)\\
	
	\While{$\mathcal{C}$.optimized is False}{
    	\ForEach{q $\gets$ Q}{
	        \uIf{$opt$ is gradient-based}{
	            $\delta$q $\leftarrow$ Compute $\nabla_q\ \min{\textbf{\text{Energy}}(\mathcal{C}, \vec{\theta}, q)}$ using VQE-based routine.\\
	        }
	        \uElseIf{$opt$ is hessian-based}{
	            $\delta$q $\leftarrow$ Compute $\frac{\nabla_q\ \min{\textbf{\text{Energy}}(\mathcal{C}, \vec{\theta}, q)}}{\nabla_q^2\ \min{\textbf{\text{Energy}}(\mathcal{C}, \vec{\theta}, q)}}$ using VQE-based routine.\\
	        }
	        \Else{
	            $opt$ is not valid. \textbf{error} 
	        }
	        
	        delQ[Q.index(q)] $\leftarrow$ $\gamma \delta$q
	   }
	   \eIf{$||delQ||_2 \geq$ ctol}{
	        \ForEach{(q, $\delta$q) $\gets$ (Q, delQ)}{
	            q $\gets$ q - $\delta$q 
	        }
	   }{
	        $\mathcal{C}$.optimized $\gets$ True\\
	        $\mathcal{C}$.molecule $\gets$ Q.values
	   }
	}
    \caption{Minimum energy configuration search for chemical system $\mathcal{C}$}
\label{algo:geom-opt}
\end{algorithm}

\subsection{\label{subsec:beyond-ground-state} Derivatives beyond ground state}

Our methodology for using VQE-based strategy for calculating energy derivatives is not limited to just ground states, but can also be adapted to excited states. Amongst the various available extensions of VQE for computing excited states for a given chemical system $\mathcal{C}$ represented by the molecular Hamiltonian $\mathcal{H}$, we base our approach on the SS-VQE protocol proposed by \citet{PhysRevResearch.1.033062}. We begin by preparing a set of $k$ mutually orthogonal quantum states, $\ket{\psi_0}, \ldots, \ket{\psi_{k-1}}$. Then, for each $i$, we calculate the expectation value $\bra{\psi^\prime_i(\theta)} \mathcal{H}(\vec{\eta}) \ket{\psi^\prime_i(\theta)}$, where $\ket{\psi^\prime_i(\theta)} = U(\vec{\theta})\ket{\psi_{i}}$. We use the weighted sum of expectation values with decreasing positive weights $w_i$ to determine the cost function $L(\vec{\theta}, \vec{\eta}) = \sum_{i}w_i \bra{\psi^\prime_i(\theta)} \mathcal{H}(\vec{\eta}) \ket{\psi^\prime_i(\theta)}$, where $w_0>w_1>\ldots>w_{k-1}>0$. 

Using an optimization routine, for a given value of system parameters $\vec{\eta}$ we find optimal parameters $\theta^*$ which minimizes the cost function $L(\vec{\theta}, \vec{\eta})$. In \cite{PhysRevResearch.1.033062}, it is shown that $\{\ket{\psi^\prime_i(\theta^*)}\}$ and $\{\bra{\psi^\prime_i(\theta^*)} \mathcal{H}(\vec{\eta}) \ket{\psi^\prime_i(\theta^*)}\}$ become approximate eigenstates and eigenvalues of $\mathcal{H}$ at the minimum value of $L(\vec{\theta}, \vec{\eta})$. Therefore, the optimal parameters $\theta^*$ give us state preparation circuits $\{U(\vec{\theta^*})\ket{\psi_{i}}\}$ for the ground state and $k-1$ excited states. Out of these many excited states, we keep only those states which preserve number of particles $\langle\mathcal{N}\rangle$ with respect to the ground state, i.e., $\bra{\psi^\prime_i(\theta^*)} \mathcal{N} \ket{\psi^\prime_i(\theta^*)} = \bra{\psi^\prime_0(\theta^*)} \mathcal{N} \ket{\psi^\prime_0(\theta^*)}$, where $i>0$ represents eigenstates other than the ground state.

One major advantage of using such a protocol is that, unlike many other protocols, this neither involves generating a new Hamiltonian by including overlaps of eigenstates nor involves diagonalization of the Hamiltonian within a subspace spanned by a chosen set of states. This means to compute the derivatives, we can still utilize the strategy as described in Section \ref{subsec:param-beyond} as we did not transform our Hamiltonian $\mathcal{H}$ to find the excited energy states. This gives the following two generalizations for first-order and seconder-order energy derivatives:

\begin{equation}\label{eq:18}
    \frac{\partial E_k^*(\vec{\eta})}{\partial \eta_j} = \bra{\psi^\prime_k(\vec{\theta}^{*})}\frac{\partial \mathcal{H}(\vec{\eta})}{\partial \eta_j}\ket{\psi^\prime_k(\vec{\theta}^{*})}
\end{equation}

\begin{equation}\label{eq:19}
\begin{split}
    \frac{\partial^2 E_k^*(\vec{\eta})}{\partial \eta_{i}\eta_{j}} =& \frac{\partial}{\partial \eta_{i}}  \bra{\psi^\prime_k(\vec{\theta}^{*})}\frac{\partial \mathcal{H}(\vec{\eta})}{\partial \eta_i}\ket{\psi^\prime_k(\vec{\theta}^{*})} \\=&  \bra{\psi^\prime_k(\vec{\theta}^{*})}\frac{\partial \mathcal{H}(\vec{\eta})}{\partial \eta_i\partial \eta_j}\ket{\psi^\prime_k(\vec{\theta}^{*})} +  \\
    \frac{2}{d\eta_i}\textbf{Re}&\Bigg[ \bra{\psi^\prime_k(\vec{\theta}^{*})}\frac{\partial \mathcal{H}(\vec{\eta})}{\partial \eta_{j}} \ket{\psi^\prime_k(\vec{\theta}^{*})_{\eta_i+d\eta_i}} -\\& \bra{\psi^\prime_k(\vec{\theta}^{*})}\frac{\partial \mathcal{H}(\vec{\eta})}{\partial \eta_{j}} \ket{\psi^\prime_k(\vec{\theta}^{*})} \Bigg]
\end{split}    
\end{equation}

\section{\label{sec:results}Results}

In this section, we describe the applications of first-order and second-order energy derivatives for the H$_2$ molecule and H$ + $H$_2 \leftrightarrow$ H$_2 + $H reaction in the minimal basis set STO-3G. These applications include: (i) obtaining minimum energy configuration, (ii) calculation of molecular response properties, and (iii) determination of transition state. Additionally, we also show the first-order excited state energy derivatives for H$_2$ molecule. The corresponding classical calculations and generation of Hamiltonian terms were done using PySCF \cite{psi4}, Gaussian \cite{g16} and OpenFermion \cite{mcclean_rubin_2020}. Their derivatives $\partial_{\eta_{i}}\mathcal{H}(\vec{\eta})$, and $\partial_{\eta_{i}} \partial_{\eta_{j}} \mathcal{H}(\vec{\eta})$ are calculated using central-differencing method with step-size $0.001$. The experiments were performed using IBM Qiskit \cite{comp_qiskit}. For noisy simulations, the noise data was taken from the IBMQ vigo backend \cite{IBMQ}, whereas Ignis and Mitiq \cite{larose2020mitiq} frameworks were used for software-level error mitigation.

\begin{figure*}[tp]
    \centering
    \begin{subfigure}[b]{0.48\linewidth}
    \begin{minipage}{.1\textwidth}
        \caption{}
        \label{fig:geom-opt-grad}
    \end{minipage}%
    \begin{minipage}{0.9\textwidth}
        \includegraphics[width=.98\linewidth]{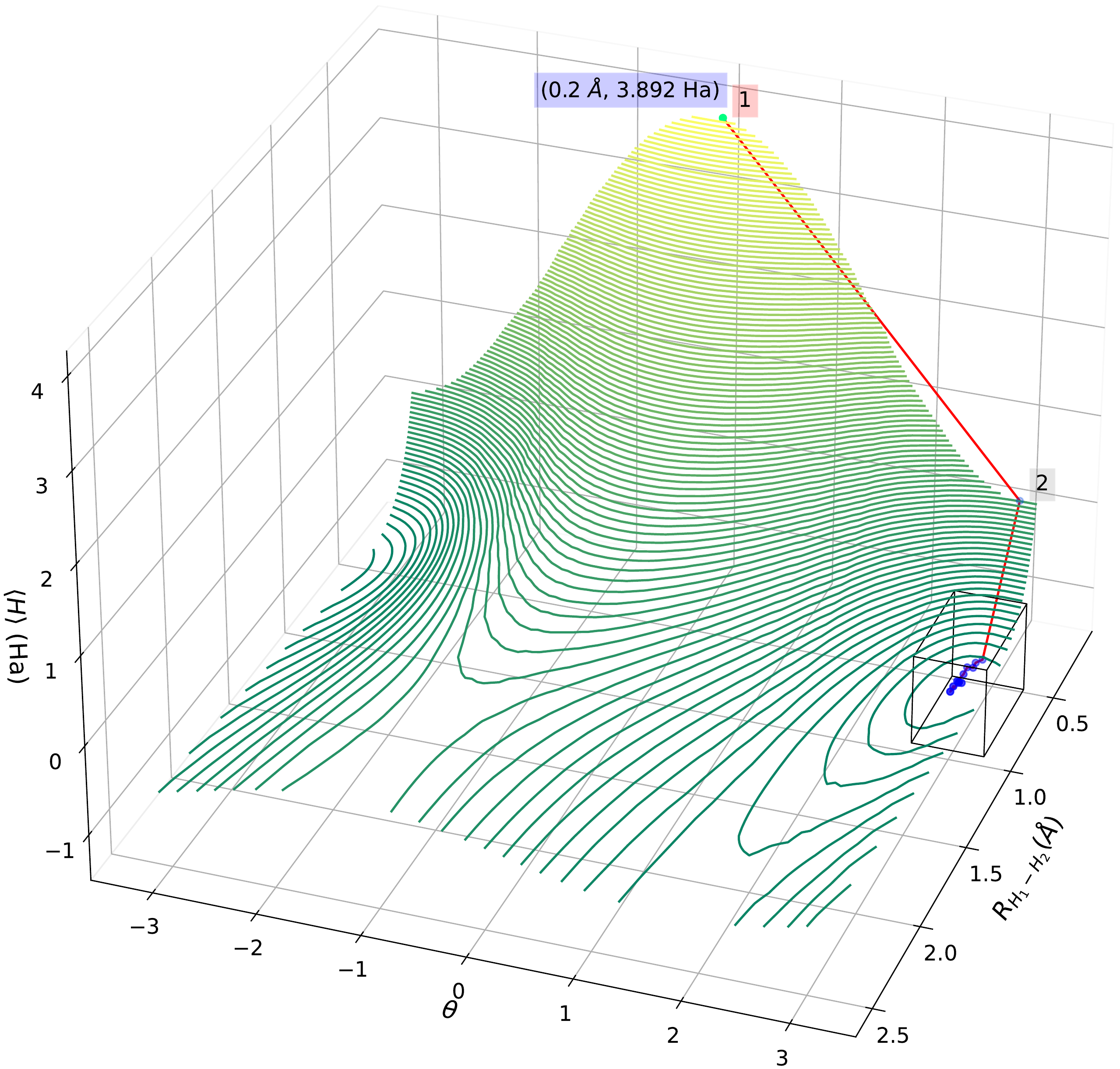}
    \end{minipage}
    \end{subfigure}
    \begin{subfigure}[b]{0.48\linewidth}
    \begin{minipage}{.1\textwidth}
        \caption{}
        \label{fig:geom-opt-gradz}
    \end{minipage}%
    \begin{minipage}{0.9\textwidth}
        \includegraphics[width=.98\linewidth]{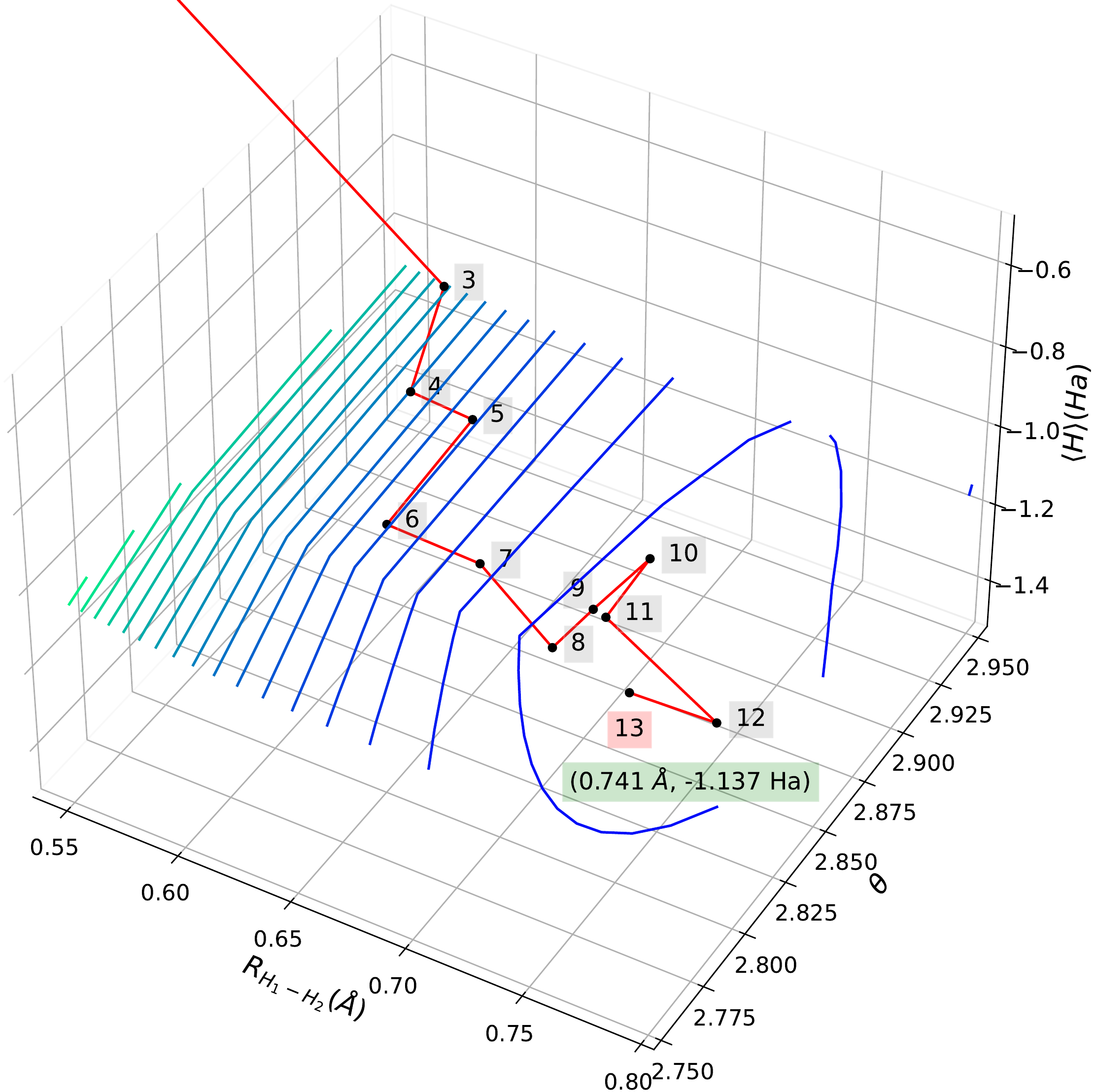}
    \end{minipage}
    \end{subfigure}
    \begin{subfigure}[b]{0.48\linewidth}
    \begin{minipage}{.1\textwidth}
        \caption{}
        \label{fig:geom-opt-hess}
    \end{minipage}%
    \begin{minipage}{0.9\textwidth}
        \includegraphics[width=.98\linewidth]{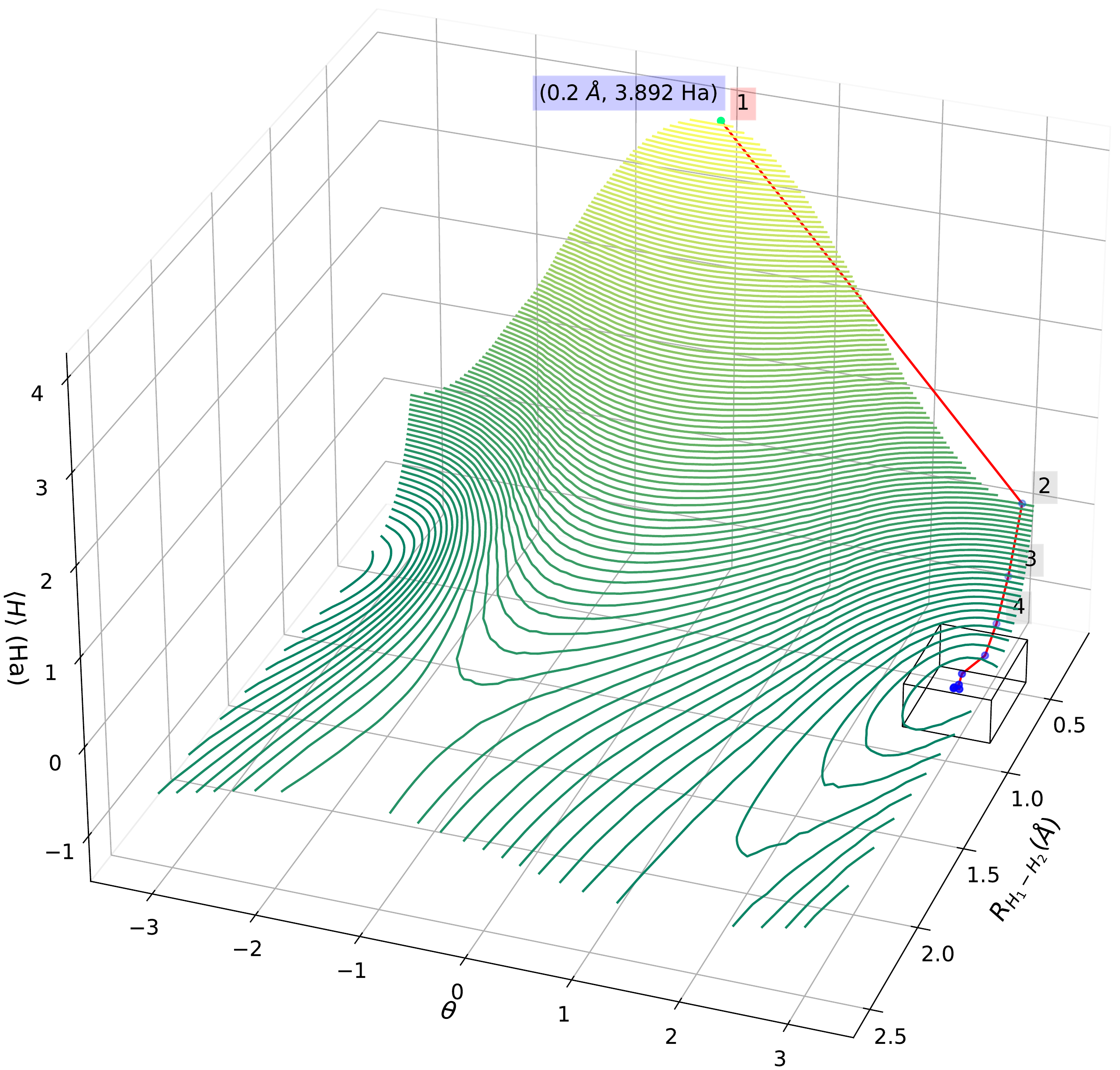}
    \end{minipage}
    \end{subfigure}
    \begin{subfigure}[b]{0.48\linewidth}
    \begin{minipage}{.1\textwidth}
        \caption{}
        \label{fig:geom-opt-hessz}
    \end{minipage}%
    \begin{minipage}{0.9\textwidth}
        \includegraphics[width=.98\linewidth]{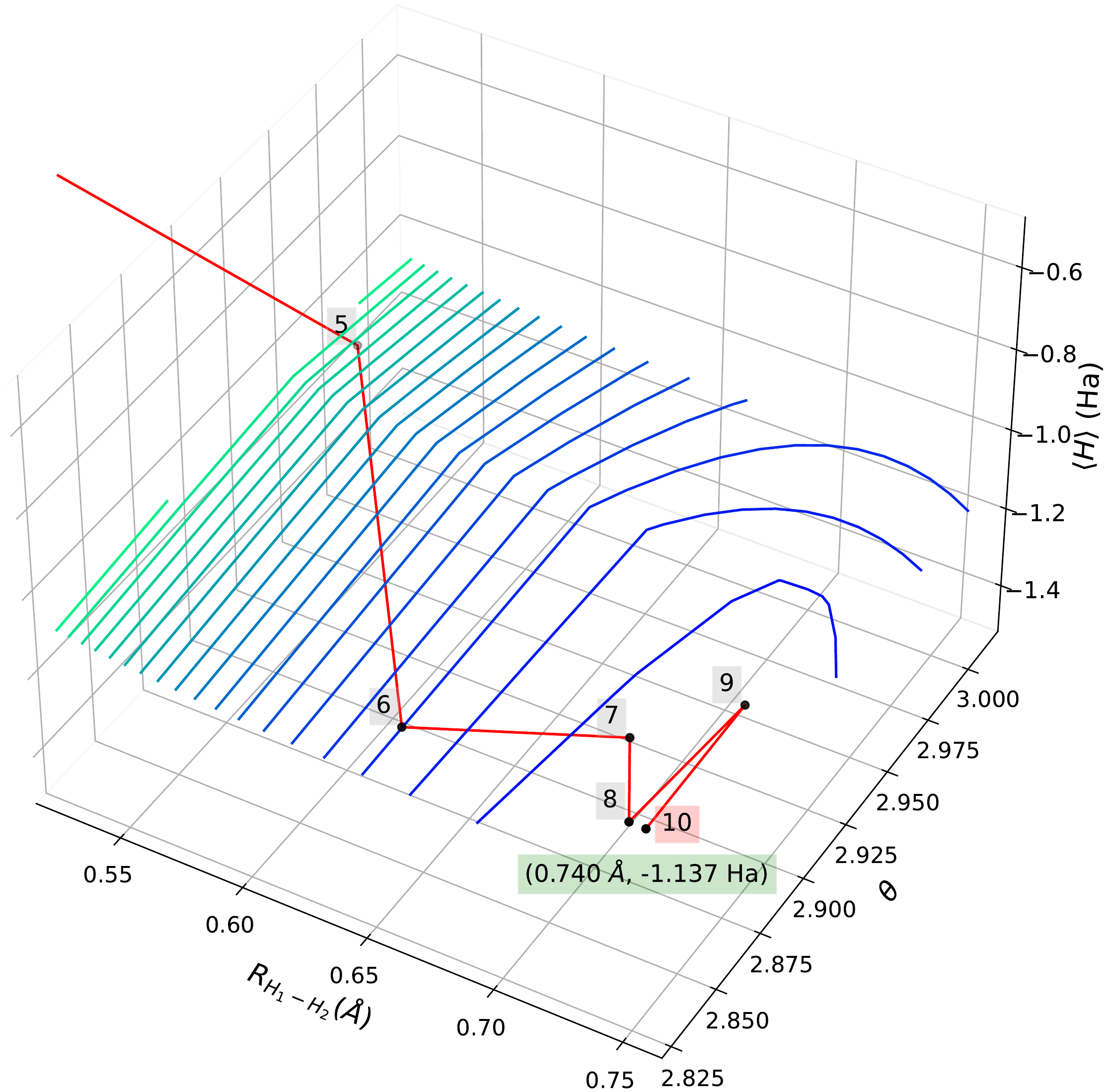}
    \end{minipage}
    \end{subfigure}
    \caption{\textbf{Minimum energy configuration search for H$_2$ molecule}: The energy surface has been calculated by evaluating $\langle \mathcal{H}(R, \theta) \rangle$ using a VQE routine for the bond lengths $R_{\text{H}_1-\text{H}_2} \in$ [0.2, 2.5] \AA, and the VQE paramater $\theta \in$ [-$\pi$, $\pi$] (Fig. \ref{fig:vqe-circuit-tapered}). Starting from an initial configuration (0.2 \AA, 3.892 Ha) on the top (maxima) of the energy surface the minimum energy configuration search was performed using gradient descent (Fig. \ref{fig:geom-opt-grad}) and Newton's method (Fig. \ref{fig:geom-opt-hess}). The former method utilizes only first-order energy derivatives (energy gradients), whereas the latter utilizes both first and second-order energy derivatives (energy gradients and hessians). The Fig. \ref{fig:geom-opt-gradz} and \ref{fig:geom-opt-hessz} represent a zoomed out view of the region bounded by the cuboids in the Fig. \ref{fig:geom-opt-grad} and \ref{fig:geom-opt-hess}. Both of the methods converged to similar optimized configurations (i) (0.741 \AA, -1.137 Ha) and (ii) (0.740 \AA, -1.137 Ha) in (i) 12 and (ii) 9 iterations respectively.} 
    \label{fig:geom-opt}
\end{figure*}

\subsection{\label{subsec:ground-state}Ground State Energy}
As explained in Section \ref{sec:vqe}, the most primitive task variational quantum eigensolvers (VQE) are designed to do is the estimation of ground-state energies for a given molecular Hamiltonian. For H$_2$ molecule, we compute ground state energies for a range of bond length $R_{\text{H}_1-\text{H}_2} \in [0.2, 1.5]$ \AA. For any given bond length $R_{i}$, we generate the molecular Hamiltonian $\mathcal{H}$ and convert it into its qubit equivalent via Bravyi-Kitaev mapping \cite{tranter_love_mintert_coveney_2018}.

\begin{equation}\label{eq:20}
\begin{split}
    \mathcal{H}^\text{BK} =& f_{0}\ \mathrm{I} + f_1\ \mathrm{Z}_0 + f_2\ \mathrm{Z}_1 + f_3\ \mathrm{Z}_2 + f_4\ \mathrm{Z}_1\mathrm{Z}_0 + \\& f_5\ \mathrm{Z}_2\mathrm{Z}_0 + f_6\ \mathrm{Z}_3\mathrm{Z}_1 + f_7\ \mathrm{X}_2\mathrm{Z}_1\mathrm{X}_0 + f_8\ \mathrm{Y}_2\mathrm{Z}_1\mathrm{Z}_0 + \\& f_{9}\ \mathrm{Z}_2\mathrm{Z}_1\mathrm{Z}_0 + f_{10}\ \mathrm{Z}_3\mathrm{Z}_2\mathrm{Z}_1 + f_{11}\ \mathrm{Z}_3\mathrm{Z}_2\mathrm{Z}_0 + \\& f_{12}\ \mathrm{Z}_3\mathrm{X}_2\mathrm{Z}_1\mathrm{X}_0 + f_{13}\ \mathrm{Z}_3\mathrm{Y}_2\mathrm{Z}_1\mathrm{Y}_0 + f_{14}\ \mathrm{Z}_3\mathrm{Z}_2\mathrm{Z}_1\mathrm{Z}_0
\end{split}
\end{equation}

\begin{figure*}[t]
    \centering
    \begin{subfigure}[b]{0.48\linewidth}
    \begin{minipage}{.1\textwidth}
        \caption{}
        \label{fig:vqe-dipole}
    \end{minipage}%
    \begin{minipage}{0.90\textwidth}
        \includegraphics[width=.98\linewidth]{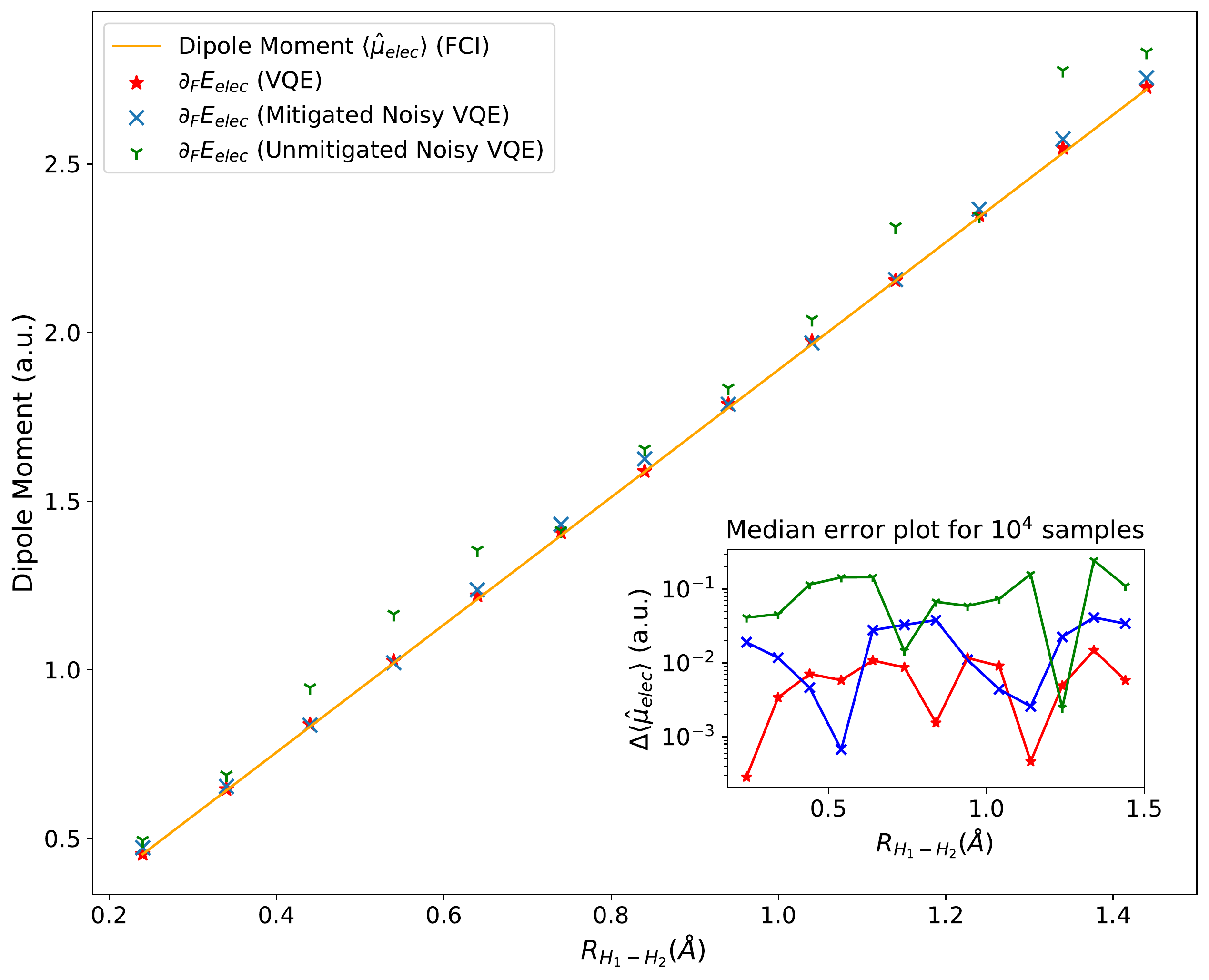}
    \end{minipage}
    \end{subfigure}
    \begin{subfigure}[b]{0.48\linewidth}
    \begin{minipage}{.1\textwidth}
        \caption{}
        \label{fig:vqe-polar}
    \end{minipage}%
    \begin{minipage}{0.90\textwidth}
        \includegraphics[width=.98\linewidth]{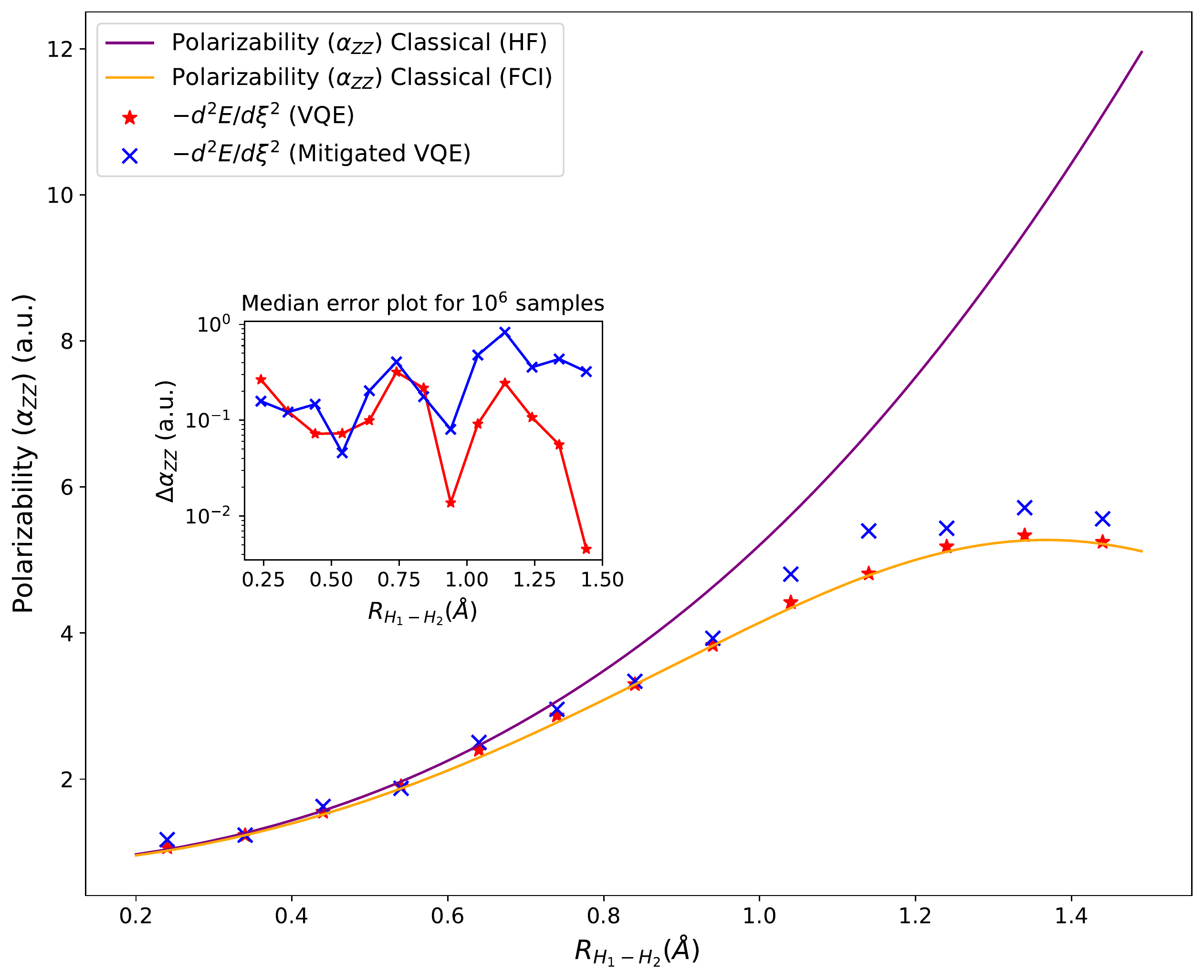}
    \end{minipage}
    \end{subfigure}
	\caption{\textbf{Molecular response properties calculation for H$_2$ molecule}: (a) dipole moments ($\mu_Z$) and (b) polarizability ($\alpha_{ZZ}$), for the bond lengths $R_{\text{H}_1-\text{H}_2} \in$ [0.2, 1.5] \AA. The solid curves indicate classically computed values for Hartree Fock and full configuration interaction (FCI) methods. Insets represent absolute errors in atomic units.} 
    \label{fig:mol-res}
\end{figure*}

We use the hardware-efficient ansatz shown in Fig. \ref{fig:vqe-circuit-full}. The results for the experiment are shown in Fig. \ref{fig:vqe-energy} with and without incorporating error-mitigation. The results show that our ansatz is powerful enough to estimate the ground state which is in accordance with the full configuration interaction (FCI) state for all the bond lengths in both noiseless simulations and mitigation-enabled experiments. This allows us to use the same ansatz for the calculation of energy derivatives as described in sections \ref{subsec:first-order} and \ref{subsec:second-order}.

\begin{algorithm}[t]
	\SetAlgoLined
	\DontPrintSemicolon
	\SetKwFor{Loop}{loop}{}{end loop}
	\KwData{reaction $\mathcal{R}$, generalized coordinates $Q$, modes $M$, learning rate $\gamma$, optimizer $opt$, tolerance $ctol$} 
	\textbf{Initialize:} $\mathcal{R}$.reactants $\leftarrow$ (Reactants' coordinates) \\
	\nonl \hspace{45pt} $\mathcal{R}$.products $\leftarrow$ (Products' coordinates) \\
	\nonl \hspace{45pt} $\mathcal{R}$.ts $\leftarrow$ False\\
	\nonl \hspace{45pt} \textbf{tstate} $\leftarrow$ None\\
	
	\While{$\mathcal{R}$.ts is False}{
    	\ForEach{m $\gets$ M}{
    	    \textbf{tstate} $\gets$ Initial guess based on $\mathcal{R}$.reactants and $\mathcal{R}$.products.\\
    	    
    	    \textbf{tstate} $\gets$ closest \textbf{stationary} point to \textbf{tstate} along $m$ via \textbf{energy configuration search}. \\
    	    
    	    A $\gets$ $\nabla^2_{m_1}$ \textbf{Energy}(\textbf{tstate}) \\
    	    B $\gets$ $\nabla^2_{m_2}$ \textbf{Energy}(\textbf{tstate}) \\
    	    C $\gets$ $\nabla_{m_1}\nabla_{m_2}$ \textbf{Energy}(\textbf{tstate}) \\

            \eIf{$A \times B - C^2 < 0$}{
    	        $\mathcal{R}$.ts $\gets$ True \\
            }{
                \textbf{continue}\\
            }    	    
	   }
	}
    \caption{Transition state search}
\label{algo:ts-opt}
\end{algorithm}

\subsection{\label{subsec:geom-opt}Minimum Energy Configuration Search}
A straightforward application of energy derivatives is to calculate the forces acting on atoms, which can be utilized to perform a minimum energy configuration search. For this, we calculate the derivatives with respect to the nuclear coordinates $\vec{R}$, which can be can be utilized in a simple gradient-based (eq. \ref{eq:22}) or hessian-based (eq. \ref{eq:23}) optimization algorithm for finding optimal nuclear coordinates $\vec{R}^*$. 

\begin{equation}\label{eq:21}
\frac{\partial E (\vec{\theta}, \vec{R})}{\partial \vec{R}} = \frac{\partial E (\vec{\theta}, \vec{R})}{\partial X} \hat{X} + \frac{\partial E (\vec{\theta}, \vec{R})}{\partial Y}\hat{Y} + \frac{\partial E (\vec{\theta}, \vec{R})}{\partial Z} \hat{Z}
\end{equation}
\begin{equation}\label{eq:22}
\vec{R}_{k+1} = \vec{R}_k - \gamma \nabla_{\vec{R}}{E (\vec{\theta}^*, \vec{R}_k)} \quad \gamma > 0
\end{equation}
\begin{equation}\label{eq:23}
\vec{R}_{k+1} = \vec{R}_k - \gamma \nabla_{\vec{R}}{E (\vec{\theta}^*, \vec{R}_k)}/\nabla_{\vec{R}}^2{E (\vec{\theta}^*, \vec{R}_k)}
\end{equation}

We perform the minimum energy configuration search for H$_2$ using gradient-descent and newton’s method \cite{10.5555/2670022}. We use a reduced two-qubit Hamiltonian $\mathcal{H}^\text{BK}_{R}$ given in \cite{PhysRevX.8.031022} and the corresponding low-depth hardware-efficient anstaz (Fig. \ref{fig:vqe-circuit-tapered}) with single variational parameter $\theta$ that was also proposed in the same work.

\begin{equation}\label{eq:24}
\begin{split}
    \mathcal{H}^\text{BK}_{R} =& g_{0}\ \mathrm{I} + g_1\ \mathrm{Z}_0 + g_2\ \mathrm{Z}_1 + \\& g_3\ \mathrm{Z}_1\mathrm{Z}_0 + g_4\ \mathrm{X}_1\mathrm{X}_0 + g_5\ \mathrm{Y}_1\mathrm{Y}_0
\end{split}
\end{equation}

Taking cues from the symmetry, we allow the movement of two $H$ atoms along the $Z$ axis only, which leads to $\partial_X E (\vec{\theta}, \vec{R}) = \partial_Y E (\vec{\theta}, \vec{R}) = 0$, i.e., $\partial_R E (\vec{\theta}, \vec{R}) = \partial_Z E (\vec{\theta}, \vec{R}) \hat{Z}$. We start with an initial bond length $R_0$ of $0.2$ \AA\ and iteratively do the following. First, we use the VQE routine to find the optimal parameter $\theta^{*}$. Second, we calculate the first-order (eq. \ref{eq:22}) and/or second-order (eq. \ref{eq:23}) energy derivatives with respect to $R_Z$ depending on the update rule of the optimization algorithm used. This process is repeated iteratively until the convergence criteria ($|\gamma\nabla_{R_{Z}} E (\vec{\theta}, \vec{R})| < 10^{-3}$) is met, where $\gamma > 0$ is the learning rate. We showcase the results for both the methods in Fig. \ref{fig:geom-opt}, where we see that the both methods converged to similar optimized configurations (i) (0.741 \AA, -1.137 Ha) and (ii) (0.740 \AA, -1.137 Ha) in (i) 12 and (ii) 9 iterations respectively. In both these cases, the final bond length and configuration energy were in agreement with their respective FCI values (0.740 \AA, -1.137 Ha).

\subsection{\label{subsec:molecule-response}Molecular Response Properties}

We can use the Taylor series expansion of $E(\vec{F})$ about $\vec{F}=0$, to express the response of a molecule under the influence of an electric field $\vec{F}$ in its environment.
\begin{equation}\label{eq:25}
E(\vec{F}) = E(0) + \bigg(\frac{\partial E(\vec{\theta}, \vec{F})}{\partial {F}}\bigg)\vec{F} + \frac{1}{2!} \bigg(\frac{\partial^2 E(\vec{\theta}, \vec{F})}{\partial F^2}\bigg)\vec{F}^2 + \ldots
\end{equation}

The Hellman-Feynman theorem \cite{doi:10.1063/1.3266959} allows us to define the electric dipole ($\mu$) as a negative derivative of the energy of the system with respect to the field $\vec{F}$. Similarly, the higher-order energy derivatives define higher-order response properties such as polarizability ($\alpha$), hyperpolarizabilities ($\beta$), hyper-hyperpolarizabilities ($\gamma$), etc. 

\begin{equation}\label{eq:26}
\mu = -\bigg(\frac{\partial E(\vec{\theta}, \vec{F})}{\partial {F}}\bigg)_{\vec{F} = 0}
\end{equation}
\begin{equation}\label{eq:27}
\alpha = -\bigg(\frac{\partial^2 E(\vec{\theta}, \vec{F})}{\partial {F}^2}\bigg)_{\vec{F} = 0}
\end{equation}

Putting eq. \ref{eq:26} and \ref{eq:27} in eq. \ref{eq:25}, we observe that while $\mu$ is a first-rank tensor, $\alpha$ on the other hand is a second-rank tensor in Cartesian basis. For simplicity, in our calculations of electric dipole moment and polarizability, we consider a constant electric field directed in the $\hat{Z}$ direction. This means that we only focus on calculating $\mu_Z$ and the $\alpha_{ZZ}$ for H$_2$ molecule. The results for dipole moment ($\mu_Z$) and the polarizability $\alpha_{ZZ}$ are shown in Fig. \ref{fig:vqe-dipole} and Fig. \ref{fig:vqe-polar} respectively using the full Hamiltonian $\mathcal{H}^\text{BK}$ (eq. \ref{eq:20}) and the corresponding hardware-efficient ansatz (Fig. \ref{fig:vqe-circuit-full}) for $D=1$. In both cases, the obtained values were in agreement with the FCI results calculated using the following qubitized dipole moment operator:

\begin{equation}\label{eq:28}
    \begin{split}
        \hat{\mu}^\text{BK}_{R}=& g_0\ \mathrm{I} + g_1\ \mathrm{X}_0 + g_2\ \mathrm{Z}_0 + g_3\ \mathrm{X}_3 + g_4\ \mathrm{Z}_3 + \\& g_5\ \mathrm{X}_0\mathrm{Z}_1 + g_6\ \mathrm{Z}_0\mathrm{Z}_1 + g_7\ \mathrm{Z}_1\mathrm{X}_2\mathrm{Z}_3 + g_8\ \mathrm{Z}_1\mathrm{Z}_2\mathrm{Z}_3
    \end{split}
\end{equation}

We also calculated the nuclear dipole moments $(\mu_N)$ to estimate the net dipole moment ($\mu_{\text{net}} = \mu_N-\mu_E$) which as expected turned out to be zero as in the case of any A$_2$ type symmetric linear molecules. Furthermore, for polarizabilities, we also compared the values of the first-order dipole moment derivative and negative of the second-order energy derivative which came out to be equal. 

\subsection{\label{subsec:transition-state}Transition State Search}
Transition state search is similar to the minimum energy configuration search, where we aim to find the point on the potential energy surface at which the energy gradient vanishes. However, unlike the latter, the former deals with finding the saddle point instead of the minimum. This requires the second-order derivative test (eq. \ref{eq:29}) since at both saddle point and minimum the energy gradient is zero. On a potential energy surface, a saddle point can be identified as the point at which the curvature at one normal mode decreases while increasing at all other normal modes, i.e., the point is maximum in one direction and minimum in the other directions. 

\begin{equation}\label{eq:29}
	(\partial^2_{R_1}E (\vec{\theta}, \vec{R}))(\partial^2_{R_2}E (\vec{\theta}, \vec{R})) - (\partial_{R_1}\partial_{R_2}E (\vec{\theta}, \vec{R}))^2 < 0
\end{equation}

Classically, there are no generalized methods that guarantee finding the right transition state. In fact, for the majority of the methods, the key ingredient for a successful search is to have a chemical intuition regarding the transition state. Similarly, the first step in our algorithm (Algo. \ref{algo:ts-opt}) is to intuitively pick a mode by guessing the transition state’s geometry. In the next step, we attempt to find the extremum along the chosen mode. Finally, if one such extremum is encountered, we perform the second-order derivative test. Passing the test means that we have encountered the transition state. Alternatively, on its failure, we repeat the previous two steps for a new chosen mode.

\begin{figure}[t]
    \centering
    \includegraphics[width=\linewidth]{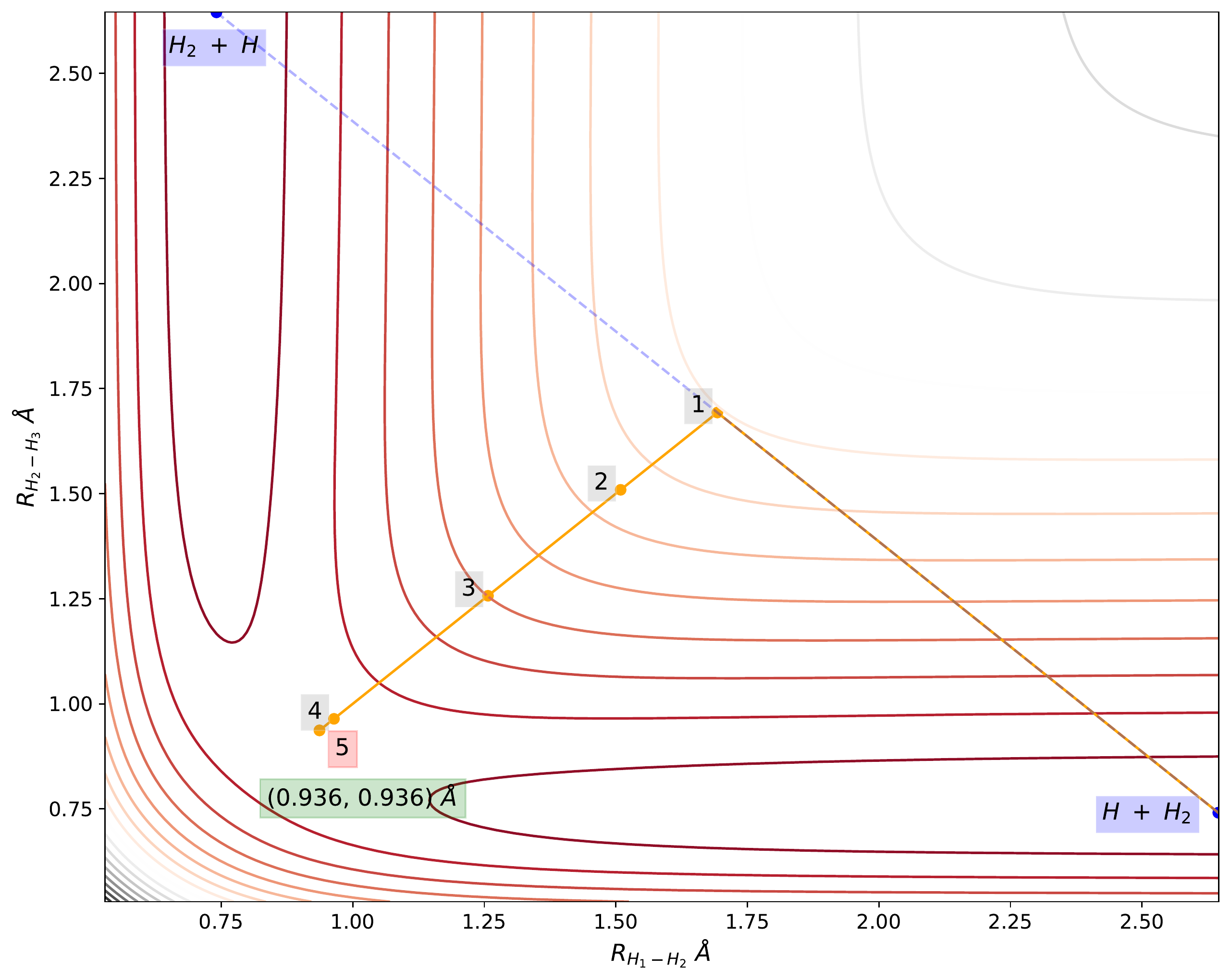}
    \caption{\textbf{Transition state search for the reaction H + H$_2$ $\leftrightarrow$ H$_2$ + H}: energy contour surface calculated by evaluating $\langle \mathcal{H}(R_1, R_2, \vec{\theta}) \rangle$ using a VQE routine for the bond lengths $R_{\text{H}_1-\text{H}_2},\ R_{\text{H}_2-\text{H}_3} \in$ [0.5, 2.65] \AA, and the VQE parameters $\theta_i \in$ [-$\pi$, $\pi$]. Starting from the reactants' configuration, an intuitive guess of transition state geometry H$_1-$H$_2-$H$_3$ is made, and a mode perpendicular to the linear path between reactants and products is chosen. At the end of five geometric optimization iterations (gray labels) extremal configuration (0.936 \AA, 0.936 \AA) is obtained, which passes the second-order test.}
    \label{fig:h3-ts}
\end{figure}

Here, we performed the transition state search for the simplest chemical reaction H$+$H$_2 \leftrightarrow$ H$_2+$H. For simplicity, we assume the reaction to be colinear and define the transition state as H$_1-$H$_2-$H$_3$, where $R_{\text{H}_1-\text{H}_2}$ and $R_{\text{H}_2-\text{H}_3}$ are the bond lengths between atom pairs H$_1-$H$_2$ and H$_2-$H$_3$ respectively. Intuitively, one can guess that the transition state should have $R_{\text{H}_1-\text{H}_2} = R_{\text{H}_2-\text{H}_3}$, and therefore in the first step we chose to search along the mode $1/\sqrt{2} (R_{\text{H}_1-\text{H}_2} - R_{\text{H}_2-\text{H}_3})$. In the second step we optimize the structure to reach an extremum according to method described in \ref{subsec:geom-opt}. Finally, we perform the second derivative test with two normal modes: (i) $1/\sqrt{2} (R_{\text{H}_1-\text{H}_2} - R_{\text{H}_2-\text{H}_3})$ and (ii) $1/\sqrt{2} (R_{\text{H}_1-\text{H}_2} + R_{\text{H}_2-\text{H}_3})$. In Fig. \ref{fig:h3-ts}, we show that we were able to successfully search the configuration of transition state H$_3$ (0.936 \AA, 0.936 \AA) in the first iteration itself with 5 iterations in the configuration optimization step. This is remarkable because the transition states can be used to determine activation energy and the reaction pathway. The former is the minimum amount of energy that must be given to enable a chemical reaction, and the latter is the steepest descent path on the potential energy surface that connects the transition state to reactants and products.

\begin{figure}[t]
    \centering
    \includegraphics[width=\linewidth]{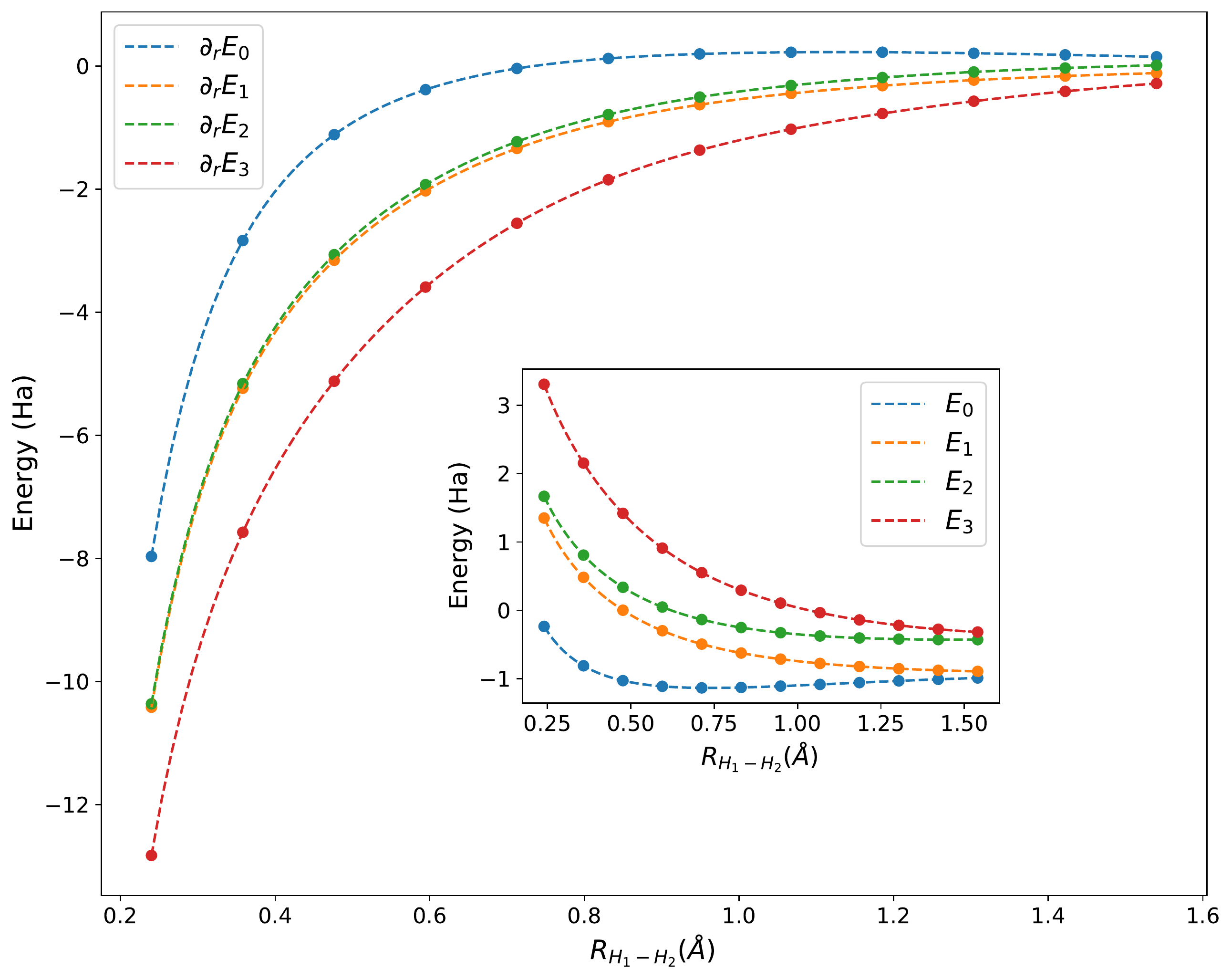}
    \caption{\textbf{Derivatives of energy states for H$_2$ molecule}: Energy derivatives of ground state and excited energy states calculated for bond length $R_{\text{H}_1-\text{H}_2} \in$ [.24, 1.54] \AA. The excited energy states shown in the inset axis have been determined by the proposed SS-VQE based protocol given in Sec. \ref{subsec:beyond-ground-state}. The dashed lines indicate the classically computed energy and energy derivative values for the full configuration method (FCI). For both energy values and energy derivative values, the optimization routine was repeated five times and the best result was chosen.} 
    \label{fig:derivative-ee}
\end{figure}

\subsection{\label{subsec:excited-state} Derivatives of Excited Energy States}

Finally, we present the results of the protocol we describe in section \ref{subsec:beyond-ground-state} for the calculation of the excited energy derivative of H$_2$ molecule in Fig \ref{fig:derivative-ee}. At a given bond length $R_{i} \in$ [.24, 1.54] \AA, we run the SS-VQE protocol with weights initialized in a geometrically decreasing manner to obtain $k$ eigenstates. Out of these, we find the valid excited states by ensuring the number of particles, $\langle \mathcal{N} \rangle$ = 2, have been conserved, where the number operator $\mathcal{N}$ is defined in the Bravyi-Kitaev formalism as:

\begin{equation}\label{eq:30}
 	\mathcal{N}^\text{BK} = 2\ \mathrm{I} - 0.5\ \mathrm{Z}_{0} - 0.5\ \mathrm{Z}_{3} - 0.5\ \mathrm{Z}_{0}\mathrm{Z}_{1} - 0.5\ \mathrm{Z}_{1}\mathrm{Z}_{2}\mathrm{Z}_{3}
\end{equation}

The potential energy curves for the first, second, and third excited energy eigenstates along with the ground state are plotted in the inset of Fig. \ref{fig:derivative-ee}. Our results from the simulation of eq. \ref{eq:18} are in good agreement with the results we generated classically.

\section{\label{sec:discussion-conclusion}Conclusion}
In this work, we have described a methodology based on the variational quantum eigensolver (VQE) for computing the derivatives of energies. The number of qubits requires by our method is equal to that of VQE. Moreover, for both ground state and excited energy states, for the calculation of single-order energy derivatives, the maximum circuit depth is equal to that of VQE, whereas, for the second-order energy derivatives, it is at most twice that of VQE for a given ansatz. Therefore, the low-depth implementation of the method makes it suitable for near-term quantum devices with software-level error mitigation techniques. We have further showcased some essential applications of the method from a quantum chemistry perspective - (i) calculation of energy derivatives with respect to some system parameters, (ii) performing the minimum energy configuration search, (iii) estimating molecular response properties such as dipole moments and polarizabilities, and (iv) finding transition state for a reaction.

Previously, a couple of works have proposed strategies in a similar spirit for evaluating energy derivatives on the quantum computer. The first work, by \citet{obrien_2019}, is based on the sum-over-state approach. Unlike ours, it involves phase estimation and eigenstate truncation instead of a hybrid variational approach. The second work, by \citet{PhysRevResearch.2.013129} is similar to ours in using a VQE-based approach. However, their method involves solving the response equation \cite{doi:10.1063/1.3266959} which requires $O(N_{\theta})$ additional measurements for the computation of derivatives of optimal variational parameters $\partial \theta_i^{*}$. Instead, our method makes use of the precomputed value of $J_2$ (eq. \ref{eq:16}) and requires just one additional VQE iteration for obtaining state preparation for $\ket{\psi(\vec{\theta}^*)}_{\eta_i + d\eta_i}$ to compute $J_1$. Besides the methodology, we also present algorithms for performing minimum energy configuration search and transition state search, where the latter has not been shown using any variational-based approach of quantum computing to the best of our knowledge.

We have shown the computational costs for our method in sections \ref{subsec:first-order} and \ref{subsec:second-order}. To compare these costs to the central differencing method, consider a variance of $\epsilon_E^2$ in estimating the energy $E(\vec{\eta})$. For this amount of variance, the numerical precision $\epsilon$ for calculating $\partial_\eta E$ and $\partial^2_\eta E$ using the centered differencing method is $O(h^2|\partial^3 E/\partial \eta_{i}^3| + \epsilon_E/h)$ and $O(h^2|\partial^4 E/\partial \eta_{i}^4| + \epsilon_E/h^2)$ respectively. To achieve similar numerical precisions from the quantum processor, one would need $O(n^4 N_\eta (\sum_P |h_P|)^2/ (h^2 (\epsilon - \lambda h^2)^2))$ measurements for first order derivatives and $O(n^4 N_\eta^2 (\sum_P |h_P|)^2/ (h^4 (\epsilon - \lambda h^2)^2))$ measurements for the second order derivatives. Here, $\lambda$ is the maximum value of $\partial^3 E/\partial \eta_i^3\ \forall\eta_i \in \eta$ in the former, whereas the maximum value of $\partial^4 E/\partial \eta_i^4\ \forall\eta_i \in \eta$ in latter. Presence of $\lambda$ in the required number of measurements makes the computation somewhat unstable when using VQE-based methods because its values is not known beforehand in both the cases. 

In the simulations and experiments for all these applications, we have used (i) simultaneous perturbation stochastic approximation (SPSA) algorithm for optimization of variational parameters $\vec{\theta}$ at each bond length, (ii) a hardware-efficient ansatz as parameterized unitary in the VQE routine, and (iii) zero noise extrapolation \cite{larose2020mitiq} along with measurement calibration to mitigate noise-induced errors. However, our method is not limited to just these choices and is further compatible with other available optimization routines \cite{PhysRevA.102.052414}, ansatz-preparation strategies \cite{2020arXiv201209265C} and error-mitigation techniques \cite{2020arXiv201008520B}. In fact, hybrid-quantum classical algorithmic approaches are, in general, are quite modular. This allows for the easy replacement of one technique present at any computational step with another one. For example, the parameter shift rule for calculating energy derivatives with respect to variational parameters can be replaced with an approach proposed by \citet{D0SC06627C}. However, the large combinations of available choices make the design process as difficult (and critical) as the hyperparameter tuning step in deep learning \cite{2015arXiv150202127C} for designing neural networks. In particular, having an intuition of the problem at hand or a well-defined search strategy \cite{2020arXiv201008561Z} plays a vital role in making these choices. For example, the number of qubits required to simulate a molecule can be reduced by identifying symmetries and freezing the orbitals that do not contribute to its evolution. 

Therefore, we conclude that a method like ours that uses depth-limited quantum circuits in a variational setting and is also compatible with techniques that reduce the need for quantum resources and mitigate errors could be used for performing tasks requiring energy derivatives on the near term quantum computers.

\section*{Acknowledgements}
We acknowledge the use of IBM Qiskit framework for this work. The views expressed are those of the authors, and do not reflect the official policy or position of IBM or the IBM Quantum team.

\bibliographystyle{apsrev4-2}

\bibliography{qchem}

\end{document}